\documentclass[conference]{IEEEtran}
\usepackage{blindtext, graphicx}
\usepackage[hyphens]{url}
\ifCLASSINFOpdf
\else
\fi

\usepackage{hhline}
\usepackage{amsmath}
\usepackage{multirow}
\usepackage{multicol}
\usepackage{booktabs} 
\usepackage{mathptmx}
\usepackage{amsfonts}
\usepackage{amssymb}
\usepackage{graphicx}
\usepackage[table,xcdraw]{xcolor}
\usepackage[utf8]{inputenc}

\begin{document}
%

\title{DroidBugs: An Android Benchmark for Automated Program Repair}


\author{
\IEEEauthorblockN{Larissa Azevedo}
\IEEEauthorblockA{Instituto de Informática\\
Universidade Federal de Goiás\\
Goiânia, Brazil\\
Email: lahazevedo07@gmail.com}
\and
\IEEEauthorblockN{Altino Dantas}
\IEEEauthorblockA{Instituto de Informática\\
Universidade Federal de Goiás\\
Goiânia, Brazil\\
Email: altinoneto@inf.ufg.br}
\and
\IEEEauthorblockN{Celso G. Camilo-Junior}
\IEEEauthorblockA{Instituto de Informática\\
Universidade Federal de Goiás\\
Goiânia, Brazil\\
Email: celso@inf.ufg.br}
}


%


\maketitle

\begin{abstract}

Automated Program Repair (APR) is an emerging research field. Many APR techniques, for different programming language and platforms, have been proposed and evaluated on several Benchmarks. However, for our best knowledge, there not exists a well-defined benchmark based on mobile projects, consequently, there is a gap to leverage APR methods for mobile development. Therefore, regarding the amount of Android Applications around the world, we present DroidBugs, an introductory benchmark based on the analyzes of 360 open projects for Android, each of them with more than 5,000 downloads. From five applications, DroidBugs contains 13 single-bugs classified by the type of test that exposed them. By using an APR tool, called Astor4Android, and two common Fault Localization strategy, it was observed how challenging is to find and fix mobile bugs.  

\end{abstract}


%
\IEEEpeerreviewmaketitle

\section{Introduction}

Despite a significant amount of the resources used in software life cycle are applied to maintenance and evolution tasks, many software still released with errors, resulting in various consequences \cite{Tricentis2017}. In this context, the field called Automated Program Repair (APR) arises to reduce maintenance costs and improve software quality. By applying diverse computational techniques, APR aiming to automatically modify a buggy program trying to fix one or more bugs.

In general, Computer Science researches employ Benchmarks to investigation different approaches on solving problems and deal with incomparable or not reproducible results. As a coming up research field, APR requires not only efforts on proposing new methods but also creating Benchmarks to evaluate the quality of new strategies and grant consistent parameters for comparisons between them \cite{7153570}.

Although few years of racing, APR field has grown rapidly then there are several Benchmarks specifically developed to evaluate APR methods. Such data typically provide a set of software projects, for one or more programming, containing buggy and fixed versions of the programs, and a certain categorization to the bugs. For example Codeflaws \cite{7965296}, DBGBench \cite{Bohme:2017:BFE:3106237.3106255}, IntroClass and ManyBugs \cite{7153570} are Benchmarks for C or C++, meanwhile Defects4J \cite{Just:2014:DDE:2610384.2628055} is for Java, and, QuixBugs \cite{Lin:2017:QMP:3135932.3135941} contains Java and Python projects.

All previous Benchmarks are based in and intended to help development of software for PC platform. However, the usage of smartphones has increased in last years, archiving 2.7 billion users, being Android the Operating System running in 77\% of all mobile devices. With about 3.3 million apps available at the Android store, the number of downloads for such a mobile OS is higher than software for PC. Hence, applications on Android must be also maintained by Automated Program Repair approaches.

Therefore, considering the aforementioned scenario, we introduce the DroidBugs a Benchmark with real and reproducible bugs, collected from five open source mobile projects, allowing to evaluate and improve APR techniques developed for repairing Android applications. DroidBugs presents 13 single-bugs, 7 revealed by instrumentation tests and 6 by local unit tests, the local of the bug, the buggy and fixed versions, and the test suite. 

The contributions of this paper may be summarized as:
\begin{itemize}
    \item Introduce and provide the first public Benchmark for APR in the context of Android development;
    \item Apply a recent APR tool, Astor4Android\footnote{https://github.com/I4Soft/Astor4Android}, and report results on localizing and repairing the bugs in the Benchmark;
    \item Point out important challenges to produce a Benchmark for mobile APR based on open source projects. 
\end{itemize}

The rest of the paper is organized as follows: the methodology of the study is described in \ref{sec:methodology}, the description of the Benchmark is presented in \ref{sec:benchmark}, an evaluation is reported in section \ref{sec:evaluation}, a discussion is conducted in \ref{sec:discussion} and, finally, conclusions are in section \ref{sec:remarks}.

\section{Methodology}
\label{sec:methodology}

This section aiming to describe the procedures used to compound DroidBugs as well as justify some decisions. The consequences of the choices are discussed later in section \ref{sec:discussion}.   

\subsection{Selecting projects}

To select projects to include in DroidBugs, we considered ones listed in F-Droid\footnote{https://f-droid.org/en}, a catalogue of FOSS (Free and Open Source Software). It was filtered only the projects available on GitHub which is the largest code host and versioning control system \cite{Gousios:2014:LGG:2597073.2597126}. Then, the projects were ranked by their popularity. Such a popularity is measured by its number of downloads in Google Play (Google online store to provide Android applications), followed by the number of commits on the respective project in GitHub.

For guaranteeing a certain relevance, 360 applications that had at least 5,000 downloads were selected. After, 50 applications were selected by applying as a criterion the use of JUnit and ones with a test suite compound at least by 10 test cases, in order to ensure that each project has a reasonable set of test cases. JUnit is a unit testing framework for Java, being the external library more included in GitHub projects written in this language \cite{misc:overops}. 

\subsection{Collecting project versions}

As Git is the versioning control system used in GitHub, a Shell script containing Git commands was run on each selected project to separate the versions in which the result of the test cases presented changes, between two consecutive versions, that typically indicate a fix of a bug. It was considered only single-bugs, that is, single defects, revealed by instrumentation test cases, unit tests run on an Android device or emulator, or by local test cases, i.e., unit tests that run only on the local machine, without real or simulated Android device.

The previous idea was implemented as follows. After cloning the project's repository from GitHub, the script filters over commits those that refer to the words ``bug'' and ``fix'', concomitantly, in their message describing the modifications made, that is, in commit message. Each separated commit was named ``PRJ\_Fix'', because it brings the developer's fix to a given error, and the version immediately preceding the fix, the parent, was called ``PRJ\_Bug'' due to it may contain the supposed bug.

\subsection{Test cases execution}

Still using Shell scripts, the test cases were executed in both PRJ\_Bug and PRJ\_Fix versions, selecting as bug those in which at least one test case negative in PRJ\_Bug becomes positive in PRJ\_Fix, indicating that a defect has been fixed by developers changes. We used the PRJ\_Fix's test suite in both versions because it is supposed to be more complete.

To run the test cases, we used the appropriate commands from Gradle, the standard compilation automation system for the Android Studio, an environment for developing Android Apps. For the instrumentation tests, an Android Emulator was created through AVD Manager. The emulator was initialized without graphical interface, to efficiency concerns. First, the test suite was run on PRJ\_Bug, and, whether there was a negative test case, then the tests were applied on PRJ\_Fix.

\section{The benchmark: DroidBugs}
\label{sec:benchmark}

After collecting the versions through the tests execution, 18 bugs were obtained from 8 applications. At this point, it was decided to consider only the bugs whose project would be executed in Astor4Android \cite{Kayque2018}. Astor4Android is an adaptation of the original library Astor to work with Android applications. Astor is an automatic program repair library  for Java with generate-and-validate techniques \cite{martinez:hal-01321615}. For our best knowledge, such a adaptation is the first and unique open source tool for Automated Program Repair for mobile development. Therefore, the initial version of the DroidBugs Benchmark is compound of 13 bugs from 5 projects (Wikipedia, Kore, Poet-Assistant, Habit and K9).

Table \ref{tb:data} summarizes the projects in DroidBugs showing their category in Google Store, number of downloads, lines of code and number of test cases available. 

\begin{table*}[ht]
\caption{\label{tb:data} Projects selected to compound the benchmark}
\renewcommand{\arraystretch}{1.15}
\resizebox{\textwidth}{!}{%
\begin{tabular}{p{3cm}llllllllllll}
\toprule
\textbf{Project}            &  &  & \textbf{Category}           &  &  & \textbf{Downloads}  &  &  & \textbf{LOC}  & &  & \textbf{Test cases} \\ \midrule
\textbf{Wikipedia Android}  &  &  & Books and References        &  &  & 10,000,000 &  &  & 197,569 & &  & 446            \\
\textbf{K-9 Mail}           &  &  & Communication               &  &  & 5,000,000  &  &  & 208,785 & &  & 1351           \\
\textbf{Kore}               &  &  & Play and edit videos        &  &  & 1,000,000  &  &  & 401,950 & &  & 131            \\
\textbf{Loop Habit Tracker} &  &  & Productivity                &  &  & 1,000,000  &  &  & 103,890 & &  & 286            \\ 
\textbf{Poet Assistant}     &  &  & Books and References        &  &  & 100,000    &  &  & 82,962  & &  & 128            \\ \bottomrule 
\end{tabular}%
}
\end{table*}

As table shows, the projects included in Benchmark have at least 100,000 downloads from Google Store, denoting their relevance. Moreover, the heterogeneity, which is an important characteristic for the Benchmark efficacy, is confirmed once only two projects are from the same category. Finally, knowing the almost of APR techniques depends on test cases, the projects in Droidbugs have from 128 to 1,351 test cases.

\begin{figure}[!ht]
\centering
\includegraphics[width=.35\textwidth]{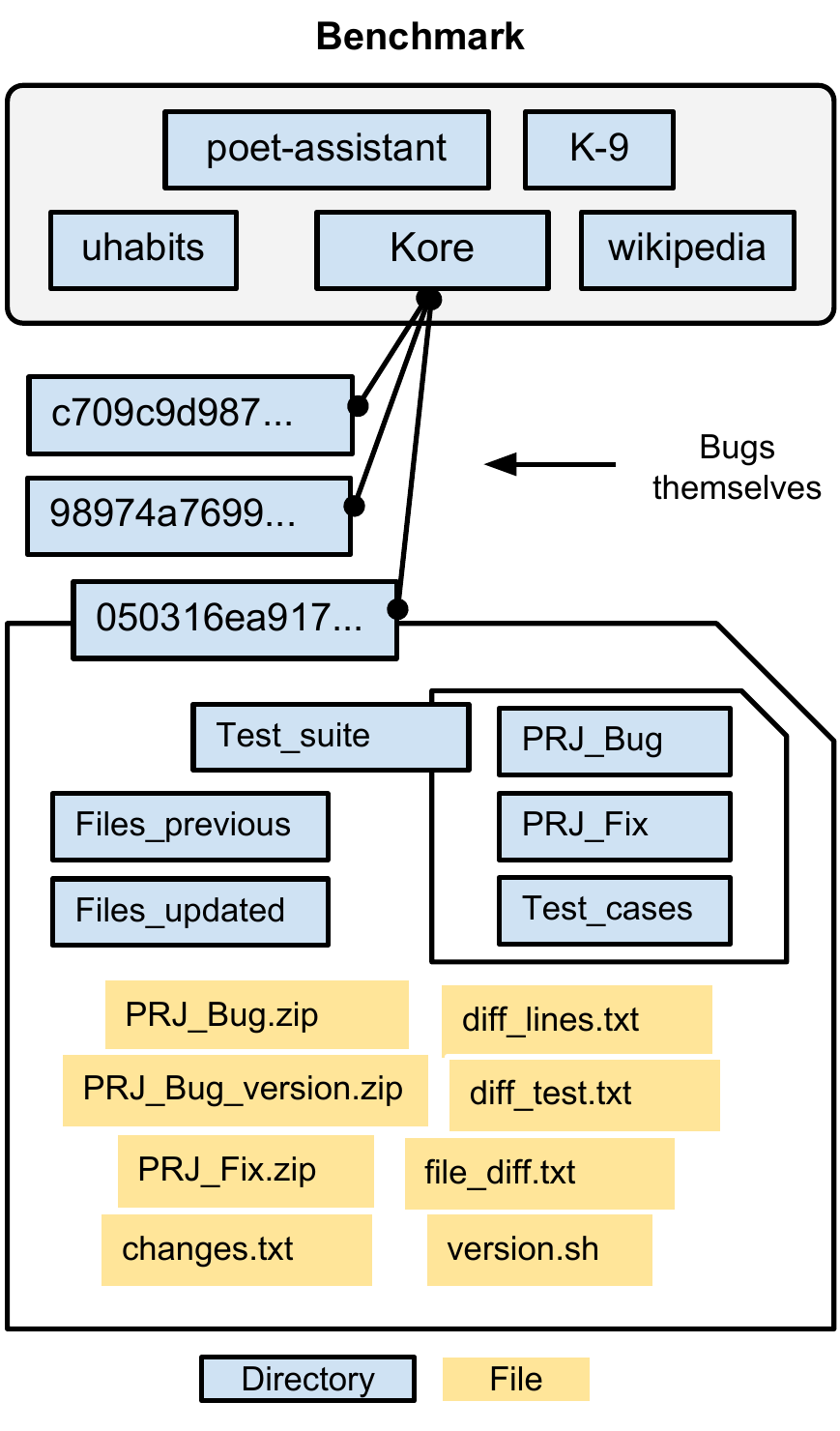}
\caption{\label{fig:illustration}Illustration of the DroidBugs Benchmark structure.}
\end{figure}

Regarding the Benchmark structure, bugs are grouped by their project and carry other relevant information inside their own directory, as shown at Figure \ref{fig:illustration}.   

In the illustration figure, one can see the Benchmark is organized first by directories for each project, inside those there are all bugs of the correspondent application. The name of each bug's directory is the identifier of the commit that repaired the respective bug. In bug's directory there are many important parts as directories and files, including compressed data.

The first group of files for each bug is compound of some zip files. \textbf{PRJ\_Bug.zip} contains the complete version in which the bug was found; \textbf{PRJ\_Fix.zip} has the version responsible to correct the bug; and, \textbf{PRJ\_Bug\_appName\_version.zip} is a version of \textbf{PRJ\_Bug} modified to able for running in Astor4Android \cite{Kayque2018}.

\textbf{Test\_suite} directory contains \textbf{Test\_Cases} in which there are all files with test cases from the fixed version; \textbf{PRJ\_Bug} has the reports from test suite execution over the buggy version; and \textbf{PRJ\_Fix} the reports from test suite execution over the fixed version. Meanwhile \textbf{Files\_previous} contains the files of the buggy version that were modified by the \textbf{PRJ\_Fix} commit, and, \textbf{Files\_uptaded} has the files modified in the fixed version commit. 

Finally, there are some txt and sh files. \textbf{changes.txt} contains the changes made in PRJ\_Bug version to enable its execution in Astor4Android, resulting in the version PRJ\_Bug\_appName\_version; \textbf{diff\_lines.txt} has the diff between the buggy and fixed version; \textbf{diff\_test.txt} has a list of test cases that changed their results in PRJ\_Fix in comparison to PRJ\_Bug; \textbf{file\_diff.txt} is a list of files modified by PRJ\_Fix; and, \textbf{appName\_version.sh} is a script to run PRJ\_Bug\_appName\_version in Astor4Android. 

For summarizing the Benchmark information, Table \ref{tb:versions} shows versions, bugs, type and number of test cases of every projects included in DroidBugs.  

\begin{table}[!ht]
\centering
\caption{\label{tb:versions}Bugs, test cases and versions of each project in DroidBugs. ``I'' and ``L'' represent Instrumentation and Local test case, respectively}
\renewcommand{\arraystretch}{1.1}
\begin{tabular}{lccc}
\toprule
\textbf{Project}            & \multicolumn{1}{l}{\textbf{Versions}} & \multicolumn{1}{l}{\textbf{Total of Bugs}} & \multicolumn{1}{l}{\textbf{Tests cases (I/L)}} \\ \midrule
\textbf{Wikipedia Android}  & 389                                           & 6                                          & 0/6                                            \\
\textbf{K-9 Mail}           & 87                                            & 3                                          & 2/1                                            \\
\textbf{Kore}               & 4                                             & 2                                          & 2/0                                            \\
\textbf{Poet Assistant}     & 12                                            & 1                                          & 1/0                                            \\
\textbf{Loop Habit Tracker} & 7                                             & 1                                          & 1/0                                            \\ \bottomrule          
\end{tabular}
\end{table}

As can be seen, the project (Wikipedia Android) with more versions (289) is also the one containing more bugs (6), and, all its bugs were exposed by local test cases. In the other hand, both Poet Assistant and Loop Habit Tracker with 12 and 7 versions, respectively, have only 1 bug, revealed by instrumentation test.

DroidBugs Benchmark is completely free and available at GitHub\footnote{\url{https://github.com/I4Soft/DroidBugs}}. We really appreciate you join us on the effort to leverage the automated program repair in mobile context.

\section{Experiments}
\label{sec:evaluation}

As stated before, the Astor4Android is the unique known tool for automatically repair a mobile application, thus, we use it to experiment the proposal Benchmark. Yet, all applications collected use Gradle as the standard compiler, but Astor4Android requires javac, then, it was created the version PRJ\_Bug\_appName\_version with the needed modifications.   

Since Astor4Android uses the negative test cases to locate errors and only considers the correctness of a project when all the test cases are positive, the test cases that have negative results when executed in PRJ\_Bug and PRJ\_Fix were removed, because they are not related to the bug fixed.

After preprocessing the project, Astor4Android performs a procedure called Fault Localization, that consists of locating the possible place of the error. In such a procedure, available test cases are executed to obtain coverage data and estimate, through a given formula, a suspicious for each lines of code, that is, a probability of the line be the cause of the bug. From lines with higher chance to contain an error, Astor4Android generates variants of the buggy application, according to a repair algorithm. When a variant compiles and allows all test cases to be positive, such a variant is considered as a fixed version. Both fault localization and repair algorithm are parameter of Astor4Android initialization.  

To find the local of errors, for each PRJ\_Bug of the 13 bugs, the tool was executed using common formulae to fault localization, Ochiai and Tarantula \cite{7985698}. The results were classified in a ranking according to the suspicion value of each line and analyzed using the wef metric, which measures the wasted effort looking for lines where there is no defect and acc@n \cite{Sohn:2017:FUC:3092703.3092717}, that counts the number of errors located in the first n positions of the ranking, for n equal to 1, 3 and 5. 

In Table 3, an identical performance of the two formulae in relation to the acc@n metric is observed. For wef, the best performance of Ochiai is influenced, mainly, by the greater number of defects not located by Tarantula. All bugs not found by both techniques are exposed by unit test case of instrumentation.

\begin{table}[!ht]
\centering
\caption{\label{tb:results}Results of fault localization with Ochiai and Tarantula over all buggy versions}
\renewcommand{\arraystretch}{1.1}
\begin{tabular}{llllll}
\toprule
\textbf{Formula}   & \textbf{acc@1} & \textbf{acc@3} & \textbf{acc@5} & \textbf{wef}         & \textbf{bugs not found} \\ \midrule
\textbf{Ochiai}     & 4     & 4     & 5     & 677.53 & 3              \\
\textbf{Tarantula} & 4     & 4     & 5     & 839.61 & 5             \\ \bottomrule
\end{tabular}%
\end{table}

Trying to fix all 13 bugs, Astor4Android was run using its three repair algorithms, JGenProg, JKali, and JMutRepair, for each bug. The first algorithm employs genetic programming to evolve individuals in order to find an individual who corrects all the defects of the under-repair App. The second tries to find a solution by removing features from the under-repair application. The last one generates mutants from the application applying pre-defined mutation operators at lines of code returned by the fault localization algorithm.

Unfortunately, none of the three APR algorithms repaired any bugs present in the Benchmark. Often, generated variants of the buggy application presented compilation errors and those that compiled did not turn all negative test cases into positive ones. These findings reinforce how hard is to automatically repair a bug in a mobile app as well as the relevance of having Benchmarks to evaluate and improve the APR techniques for this type of program.

\section{Discussion}
\label{sec:discussion}

The results obtained can be analyzed under the prism of the difficulties and challenges from the methodology adopted and issues about Automatic Program Repair for Android itself. Considering all existing Android applications, many of them were impossible to use due to some constraints defined to standardize the projects in the Benchmark. Although popularity and greater use were the criteria for choosing the required features, satisfy such criteria reduced the search space of bugs. Therefore, we believe that is necessary to loosen some constraints to increase the number of achievable bugs. 

From the millions of Android apps, only about 1,500 are open source and listed by F-Droid. Some of these projects are not hosted on GitHub or even do not apply versioning control. Most of the projects, listed in the phase ``Selecting projects'', do not have a satisfactory number of test cases or were not written in Java. Thus, only 50 applications remained to continue for the phases ``Collecting projects'' and `` Executing test cases'', and, consequently, limited the final applications in the Benchmark.

During bugs mining process, it was notice a missing evolution of the test suite. We observed for some projects, although there was a set of reasonable initial test cases, they do not evolve simultaneously to the code. Thus, there are bug fixes without addition or modification in the test suite, which makes it impossible to discover the error. It was attempted to create test cases manually from the commit message and changes introduced by it. However, this strategy was not effective in time spent, since it requires a great deal of knowledge of the application code, which is a very expensive task.

Finally, many of the filtered commits coming from the 50 selected projects do not compile in both versions PRJ\_Bug and PRJ\_Fix; or, there is the addition of new methods and/or classes in the version in which the bug was fixed. This addition generates compilation errors in PRJ\_Bug when using PRJ\_Fix test suite, because the test cases use the code snippets included in the fix commit. For up caming  versions of DroidBugs we intended to consider this kind of bug classifying them as ``omission errors''.

\section{Conclusion}
\label{sec:remarks}

Automated Program Repair (APR) is a recent research field that have gained attention in last years. As another computational areas, APR approaches requires well-defined data to compare existing ones or evaluate new methods. Unfortunately, despite the huge dissemination of mobile apps on Android platform, for our best knowledge, there is no public Benchmark to APR algorithms for mobile apps. 

Therefore, we presented DroidBugs the first public Benchmark based on open source Android projects. Composed of 5 Android applications, DroidBugs provides 13 single-bugs revealed by two type of test cases. By applying an APR tool called Astor4Android, all bugs were investigated regarding to locate and fix them. Challenges to mining the bugs and the results obtained on experimenting them may be useful for next efforts in this research field.

\section*{Acknowledgment}

Authors would like to thank the Fundação de Amparo à Pesquisa do Estado de Goiás (FAPEG) and the Programa Institucional de Bolsas de Iniciação Cientifica of UFG for suppoting this research. 

\bibliographystyle{IEEEtran}


%

%




\end{document}